\documentclass[longauth,referee]{aa} 

\usepackage{graphicx}
\usepackage{txfonts}

\begin{document} 

\title{
A companion on the planet/brown dwarf mass boundary on a wide orbit discovered by gravitational microlensing
}

\titlerunning{Planet/brown dwarf on a wide orbit}

\authorrunning{Poleski et al.}

\author{R.~Poleski\inst{1,2}, A.~Udalski\inst{2}, I.~A.~Bond\inst{3}, 
J.~P.~Beaulieu\inst{4,5,6}, C.~Clanton\inst{7}, S.~Gaudi\inst{1}, \\ 
and, \\
M.~K.~Szyma\'nski\inst{2}, I.~Soszy\'nski\inst{2}, P.~Pietrukowicz\inst{2}, 
Szymon~Koz\l{}owski\inst{2}, J.~Skowron\inst{2}, \L{}.~Wyrzykowski\inst{2}, 
K.~Ulaczyk\inst{8,2}, \\ 
and, \\
D.~P.~Bennett\inst{9},
T.~Sumi\inst{10},
D.~Suzuki\inst{9},
N.~J.~Rattenbury\inst{11},
N.~Koshimoto\inst{10},
F.~Abe\inst{12},
Y.~Asakura\inst{12},
R.~K.~Barry\inst{9},
A.~Bhattacharya\inst{9,13},
M.~Donachie\inst{11},
P.~Evans\inst{11},
A.~Fukui\inst{14},
Y.~Hirao\inst{10},
Y.~Itow\inst{12},
M.~C.~A.~Li\inst{11},
C.~H.~Ling\inst{3},
K.~Masuda\inst{12},
Y.~Matsubara\inst{12},
Y.~Muraki\inst{12},
M.~Nagakane\inst{10},
K.~Ohnishi\inst{15},
C.~Ranc\inst{9},
To.~Saito\inst{16},
A.~Sharan\inst{11}, 
D.~J.~Sullivan\inst{17},
P.~J.~Tristram\inst{18},
T.~Yamada\inst{10},
T.~Yamada\inst{19}, 
A.~Yonehara\inst{19},
\\
and, \\
V.~Batista\inst{4}, J.~B.~Marquette\inst{4}
}

\institute{Department of Astronomy, Ohio State University, 140 W. 18th Ave., Columbus, OH 43210, USA 
\\ \email{poleski.1@osu.edu}
\and Warsaw University Observatory, Al. Ujazdowskie 4, 00-478 Warszawa, Poland 
\and Institute for Natural and Mathematical Sciences, Massey University, Private Bag 102904 North Shore Mail Centre,\\Auckland 0745, New Zealand
\and Sorbonne Universit\'es, UPMC Univ Paris 6 et CNRS, UMR 7095, Institut d’Astrophysique de Paris, 98 bis bd Arago, F-75014 Paris, France
\and LESIA Observatoire de Paris, Section de Meudon 5, place Jules Janssen F-92195 Meudon, France
\and School of Physical Sciences, University of Tasmania, Private Bag 37 Hobart, Tasmania 7001 Australia
\and NASA Ames Research Center, Space Science \& Astrobiology Division, Moffett Field, CA 94035, USA
\and Department of Physics, University of Warwick, Coventry CV4 7AL, UK 
\and Laboratory for Exoplanets and Stellar Astrophysics, NASA/Goddard Space Flight Center, Greenbelt, MD 20771, USA
\and Department of Earth and Space Science, Graduate School of Science, Osaka University, 1-1 Machikaneyama, Toyonake, Osaka 560-0043, Japan
\and Department of Physics, University of Auckland, Private Bag 92019, Auckland, New Zealand
\and Institute of Space-Earth Environmental Research, Nagoya University, Furo-cho, Chikusa, Nagoya, Aichi 464-8601, Japan
\and Department of Physics, University of Notre Dame, Notre Dame, IN 46556, USA
\and Okayama Astrophysical National Astronomical Observatory, 3037-5 Honjo, Kamogata, Asakuchi, Okayama 719-0232, Japan
\and Nagano National College of Technology, Nagano 381-8550, Japan
\and Tokyo Metropolitan College of Industrial Technology, Tokyo 116-8523, Japan
\and School of Chemical and Physical Sciences, Victoria University, Wellington, New Zealand
\and University of Canterbury Mt John Observatory, PO Box 56, Lake Tekapo 7945, New Zealand
\and Department of Physics, Faculty of Science, Kyoto Sangyo University, 603-8555 Kyoto, Japan
}

\date{Received April 3, 2017; accepted May 4, 2017}
 
  \abstract{
We present the discovery of a substellar companion to the primary host lens in 
the microlensing event MOA-2012-BLG-006.  The companion-to-host mass ratio is 
$0.016$, corresponding to a companion mass of $\approx8~M_{\rm Jup} (M_*/0.5M_\odot)$.  
Thus, the companion is either a high-mass giant planet or a low-mass brown dwarf, 
depending on the mass of the primary $M_*$.  The companion signal was separated 
from the peak of the primary event by a time that was as much as four times 
longer than the event timescale.  
We therefore infer a relatively large projected separation of the companion from 
its host of  $\approx10~{\rm a.u.}(M_*/0.5M_\odot)^{1/2}$ for a wide range 
(3-7 kpc) of host star distances from the Earth.  We also challenge a previous 
claim of a planetary companion to the lens star in microlensing event OGLE-2002-BLG-045.  
}


   \maketitle

\section{Introduction} 

Brown dwarfs and planets are intrinsically faint objects and 
different detection techniques have to be used to explore a wide range of 
properties of these sub-stellar objects. 
Every detection technique has its own limitations and leads to a different 
kind of information when a new object is detected. 
Despite the large number of observational and theoretical studies 
\citep[e.g.,][]{beichman14,chauvin15,foremanmackey16,wilson16}, 
we are still far from a detailed understanding of the demographics of 
brown dwarf and planet populations that are also companions to stars. 
There is even a lack of consensus on the appropriate border line between planets 
and brown dwarfs \citep{boss03,grether06,spiegel11,chabrier14}. 
The obvious way to increase our knowledge of sub-stellar mass objects is 
by discovering more objects and, in particular, by discovering and 
characterizing objects that question our 
current understanding of planet and brown dwarf formation and evolution. 

Here, we present the discovery of a binary system MOA-2012-BLG-006L 
with a mass ratio of $0.016$ 
and projected separation of roughly $10~{\rm a.u.}$ 
Both components of the system were detected using the gravitational microlensing 
method. The advantage of this method is that it is sensitive 
to the mass of the objects, rather than their luminosity. 
As a result, microlensing enables the discovery of systems that 
are inaccessible to other techniques. 
First, the system distance of a few kpc prevents the detection of light from 
the lower-mass component via direct imaging. 
Second, the radial velocity signal of a long-period, low-mass companion 
to the faint host is out of reach of current techniques. 
Finally, the projected separation of about $10~{\rm a.u.}$ results in 
the extremely low probability of observing the transit, even if 
a population of similar systems was observed
using photometric methods. 
We note that there are three other systems that were discovered using 
microlensing and contain either brown dwarfs or planets: 
MOA-2007-BLG-197L \citep{ranc16}, 
MOA-2010-BLG-073L \citep{street13}, and 
MOA-2011-BLG-322L \citep{shvartzvald14}. 
These three systems have smaller separations and 
higher mass ratios compared to the system reported here, MOA-2012-BLG-006L. 
The distribution of mass ratios for binary lens microlensing events was 
recently investigated by \citet{shvartzvald16a}. They found that the mass 
ratio distribution shows the minimum and this minimum is close to 
the mass ratio of MOA-2012-BLG-006L ($0.016$).

Explaining the formation of the MOA-2012-BLG-006L system poses significant challenges. 
The mass of the protoplanetary disc is typically $0.002$-$0.006$ of 
the host mass \citep{andrews13}, hence, any planet that forms in 
a protoplanetary disc that follows this observational trend 
cannot have a larger mass ratio. If the protoplanetary disc in MOA-2012-BLG-006L 
had a mass ratio close to the typical values, then the lower-mass object 
should be classified as a brown dwarf. However, there is a wide range 
of measured disc masses at 
fixed host mass \citep{andrews13}. In the extreme cases, estimated 
disc masses are close to $0.2$ of the host mass \citep{andrews09}. 
In these extreme cases, the total mass of the disc is sufficient to form 
planets with mass ratios similar to MOA-2012-BLG-006Lb. 
The planetary formation scenario poses an additional question: 
how did a planet 
so massive end up on an orbit that is at least eight times larger than 
the snow line distance ($\approx1.3~{\rm a.u.}$ in this case)?  
Most massive planets formed by core accretion should do so just beyond 
the snow line, where the protoplanetary disc is still relatively dense 
and ices can condense.
Furthermore, if the planet formed via gravitational instability, 
we might expect it to be on an orbit wider by a factor of a few 
in semimajor axis 
\citep[which, depending on the projection, it may actually be; ][]{dodson-robinson09}

In the following Section we describe photometric observations 
leading to the discovery of MOA-2012-BLG-006Lb. 
In Section~\ref{sec:analysis} we analyze photometric data and derive 
the system properties using a Galactic model. 
The degeneracies in the microlensing model fitting are described in detail. 
The following section presents high-resolution observations of the event. 
Section~\ref{sec:ob02045} discusses another microlensing event 
(OGLE-2002-BLG-045) that showed a possible anomaly 
that could be fitted with a planetary model. We conclude that the anomaly was not 
real and there is no evidence for a planet. 
We end with conclusions. 

\section{Photometric observations} 
\label{sec:obs}

The microlensing event MOA-2012-BLG-006 was announced by the MOA group
\citep[Microlensing Observations in Astrophysics; ][]{bond01} on 
Feb 9, 2012 (${\rm HJD'} \equiv {\rm HJD}-2450000 = 5967.3$) at 
(R.A., Dec.) = ($18^{\rm h}01^{\rm m}46\fs31$,~$-29\degr06\arcmin31\farcs6$) 
(Galactic coordinates $l\approx1\fdg64$, $b\approx-3\fdg13$).
The event was found very early during the bulge observing season. During that
time, bulge is visible only for a short time each night from any single site. 
The chances of discovering planets so early during the bulge observing 
season are low and most of the follow-up surveys do not start their normal 
operations before about a month later. Hence, survey observations 
are the only way to find planets that show their signatures so
early in the season. 
The same event was alerted by the OGLE survey 
\citep[Optical Gravitational Lensing Experiment; ][]{udalski03} 
on Feb 13, 2012 (${\rm HJD'} = 5971$) in the first batch of 
the microlensing events in 2012 and labeled OGLE-2012-BLG-0022. 
The OGLE and the MOA survey telescopes are well separated in geographic longitude, 
which allows coverage of different parts of the light curve. 
Below we describe the datasets produced by the two surveys. 

\begin{figure}
\centering
\includegraphics[width=.8\hsize]{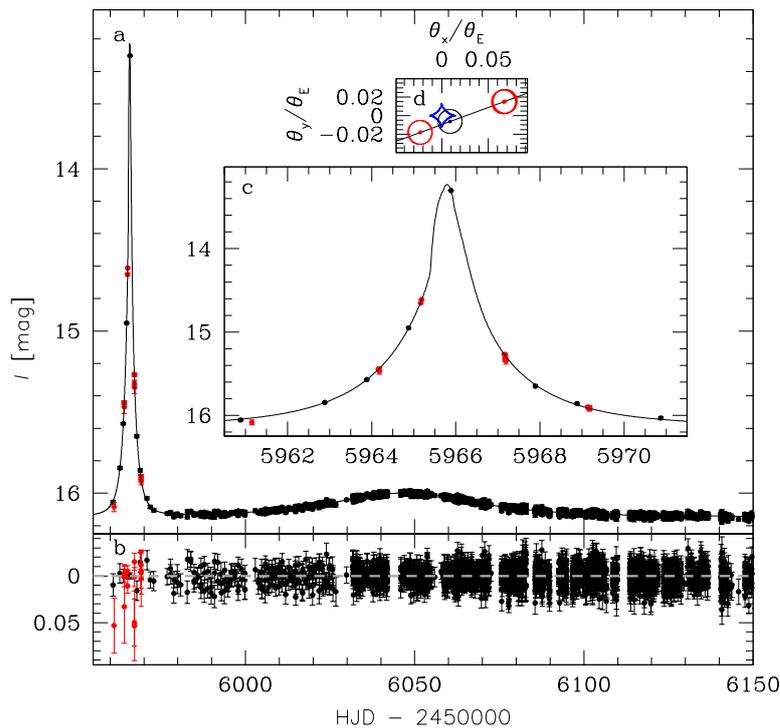}
\caption{Light curve of MOA-2012-BLG-006 = OGLE-2012-BLG-0022. 
OGLE and MOA data are marked using black and red points, respectively. 
{\it Panel a} presents the 2012 light curve with both subevents. 
{\it Panel b} shows the model residuals. 
{\it Panel c} gives zoom-in on the anomaly. 
{\it Panel d} presents the source trajectory (black line) 
relative to the planetary caustic (blue curve), 
which is at the origin of the coordinate system. The host star is located at 
$(\theta_x/\theta_{\rm E}, \theta_y/\theta_{\rm E}) = (4.17, 0.0)$. 
Source positions from one OGLE night and two MOA nights are marked 
and are aligned with photometry shown in {\it panel c}. 
The circles have a radius of $\rho$. 
}
\label{fig:lc}
\end{figure}

The main photometric dataset comes from the OGLE survey, 
which uses a 1.3m telescope located at Las Campanas Observatory (Chile). 
The telescope is equipped with the 32-CCD mosaic camera that gives a $1.4~{\rm deg^2}$ 
field of view \citep{udalski15b}. 
There are eight photometric epochs during the anomaly 
(i.e., ${\rm HJD'}$ between 5960 and 5971) and 
2306 more measurements during the 2012 bulge season -- see light curve in Figure~\ref{fig:lc}. 
We also included 813 datapoints from 2011 in order to ensure that the baseline 
brightness is correctly measured. 
The OGLE survey performs most of the observations in the $I$-band 
and we use only these data for fitting. The $V$-band data do not cover 
the anomalous part of the light curve and are only used to derive source properties. 
The photometry was performed using the Difference Image Analysis (DIA) method 
\citep{alard00,wozniak00}. The photometric uncertainties 
were corrected using the prescription 
presented by \citet{skowron16a}. There are two relatively bright 
field stars that are very close to
the event: $1.1$ and $1.4~{\rm arcsec}$ away with an $I$-band brightness of 
$17.6$ and $16.2~{\rm mag}$, respectively. 
The two stars can affect the photometry of the event. 
Indeed, we found that seeing variations marginally influence brightness 
measurements -- the target gets fainter by $0.005~{\rm mag}$ 
for an increase in seeing FWHM of 
$1~{\rm arcsec}$. This effect was subtracted from the OGLE data. 

The second dataset used comes from the MOA survey. The MOA survey operates 
a 1.8m telescope situated at Mt.~John Observatory (New Zealand). 
The filter used for observations is a custom wide-band optical filter.  
The camera consists of ten CCD detectors and gives a $2.2~\rm{deg^2}$ field 
of view \citep{sako08}. 
The MOA observing site has poorer weather and seeing conditions as 
compared to the OGLE site, but 
enables observations of microlensing events 
when Galactic bulge is invisible from Chile. 
Photometry was performed using the DIA method \citep{bond01}. 
MOA data for the analyzed event are more affected by the variable seeing and 
additionally show dependence of measured brightness on 
the airmass. 
Unfortunately, the specific way in which data 
are affected changes over time. We see these changes 
both during the event, and during the other observing seasons. 
In order not to include the affected data in the fit, we restricted the MOA dataset 
to the fourteen epochs that are closest to the anomaly, i.e.,
from ${\rm HJD'}=5961.1$ to $5969.2$. 
Similar issues with a nearby star contaminating photometry of 
the microlensing event were faced by \citet{gould14a} who analyzed 
the event OGLE-2013-BLG-0341. 
We note that the remaining MOA data will not improve accuracy of the fitted event 
properties, because with the exception of the anomaly, the event was of 
a low magnification and during bulge observing season 
the OGLE cadence of $20~{\rm min}$ is more than sufficient to characterize the light curve. 
We also checked that the other bulge photometric survey operating at that time -- 
VISTA Variables in the V\'ia L\'actea \citep[VVV;][]{minniti10} -- 
did not collect any data of this field during the anomaly.

\section{Analysis} 
\label{sec:analysis}

The light curve of MOA-2012-BLG-006 resembles a superposition of two 
point-source/point-lens microlensing events. Light curves of this type can be 
produced in two physically different scenarios \citep{gaudi98}. 
First, the lens can be a single object 
and the source can be a binary system leading to two subevents with the same 
Einstein timescale $t_{\rm E}$. We tried to fit the binary source model to the 
observed light curve, but could not find a good fit. 
After rejecting the binary source model, we are 
left with only one other possibility -- the lens is a binary system and a single 
source is magnified \citep{gaudi98}. If the two subevents are not significantly affected by the 
caustics (curves on which point source magnification is infinite), then the 
mass ratio of the two lens components is a square of the $t_{\rm E}$ ratio 
of the subevents.  
In the present case, a simple examination of the light curve by eye suggests that 
the lower mass object is either a planet or a brown dwarf if the host is 
a typical main sequence star. 
The first subevent has a higher magnification, even though it was caused by the lower 
mass object. 
The magnification of the subevent depends primarily on the impact parameter, not the lens mass. 

To fit the microlensing model we evaluated magnification using the inverse ray 
shooting method for the highest magnified points and the hexadecapole approximation
\citep{gould08,pejcha09} for the adjacent parts of the light curve. 
We used the complex polynomial root solver by \citet{skowron12}. 
Based on \citet{claret11} and source properties derived from the initially fitted model, 
we set the limb darkening coefficients to $\Gamma_I=0.502$ ($u_I=0.602$)  
and $\Gamma_{\rm MOA} = 0.588$ ($u_{\rm MOA} = 0.681$). 
We note that initial fitting was performed to the OGLE data only, but the results 
do not qualitatively differ from fits to the OGLE and MOA data. 

We first tried to fit the model using Monte Carlo Markov Chain (MCMC) that is 
typically used for the analysis of the microlensing events. 
We used MCMC implementation by \citet{dong07} and \citet{poleski14c}. 
Even though we run the MCMC multiple times with a number of settings, 
MCMC failed to produce a converging chain and 
hence we could not use it to fit a microlensing model. 
The triangle (or corner) plot showed that almost all 
two-parameter marginalized $\chi^2$ hypersurfaces had approximately ellipsoidal shapes 
but still the chain was not converging. 
The only exceptions were $\chi^2$ hypersurfaces where one of the parameters was 
the angular source radius relative to the Einstein ring radius ($\rho$). 
We fully understood the reason for failure in the MCMC runs 
only after transforming the microlensing model was 
transformed to a different set of parameters. A default set of seven parameters 
that describes a binary lens model consists of: the three point lens parameters, $\rho$, 
and the three 
parameters that describe the lens companion. The three point lens parameters are:
$t_0$ -- the epoch of minimum source-lens separation, $u_0$ -- the minimum 
separation relative to Einstein ring radius $\theta_{\rm E}$, and $t_{\rm E}$. The lens companion is 
described by: $\alpha$ -- the angle between the lens axis and the source trajectory, $s$ -- 
the separation of the lens components relative to $\theta_{\rm E}$, and $q$ -- the mass 
ratio. In addition to the binary lens parameters that completely describe 
the magnification, the model also contains source and blending
fluxes for each photometric system.  
The default binary lens parametrization is not optimal for 
fitting all the microlensing events because the parameters $\alpha$, $q$, and 
in many cases $s$ are not directly constrained by the light curves, 
in the sense that the observable properties of the microlensing event are not directly 
relatable to these parameters 
\citep{cassan08,sumi10,skowron11,kains12}. 

The event MOA-2012-BLG-006 shows two well-separated subevents and their parameters 
(maximum magnification, its epoch, and the length of the subevent) are 
well-constrained by the data, assuming the blending flux is known. 
Hence for fitting, we used, instead of the default set of parameters, 
$\rho$ and the three point lens parameters measured separately for each of the components 
($t_{\rm E}$) or relative to their caustics ($t_{0}$ and $u_{0}$). 
We note that in this two-component parameterization either $u_{0,1}$ or $u_{0,2}$ 
has to be a signed quantity in order to make a distinction between the source passing 
both caustics on same or opposite sides (unlike in a point-source/point-lens model without parallax). 
The conversion between both binary lens parameterizations is based on simple geometry and 
the equation for distance between the central and the planetary caustics: 
$s-s^{-1}$ \citep{han06}. 

\begin{figure}
\centering
\includegraphics[width=.75\hsize,bb=5 20 550 540]{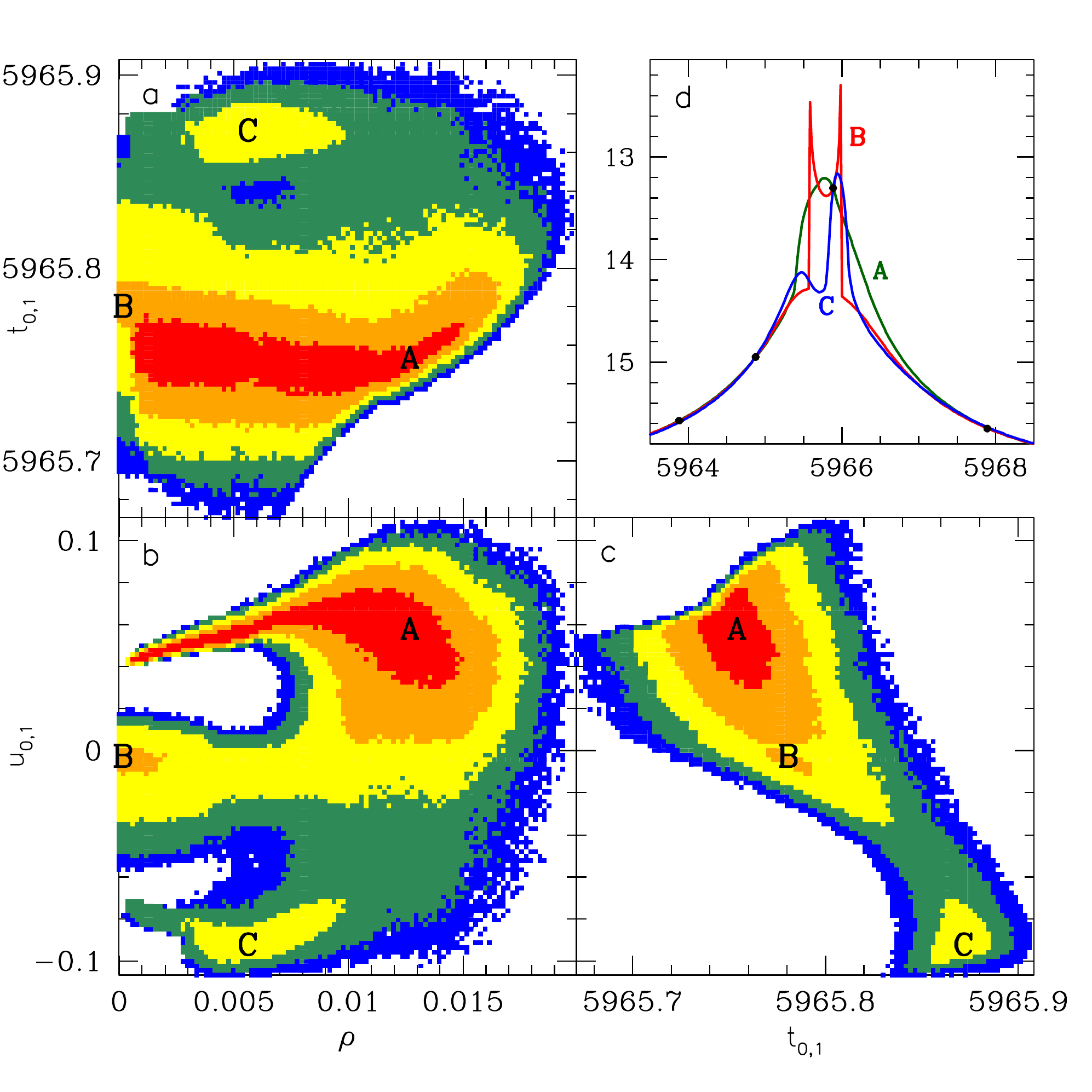} \includegraphics[width=.244\hsize,bb=15 -180 292 540]{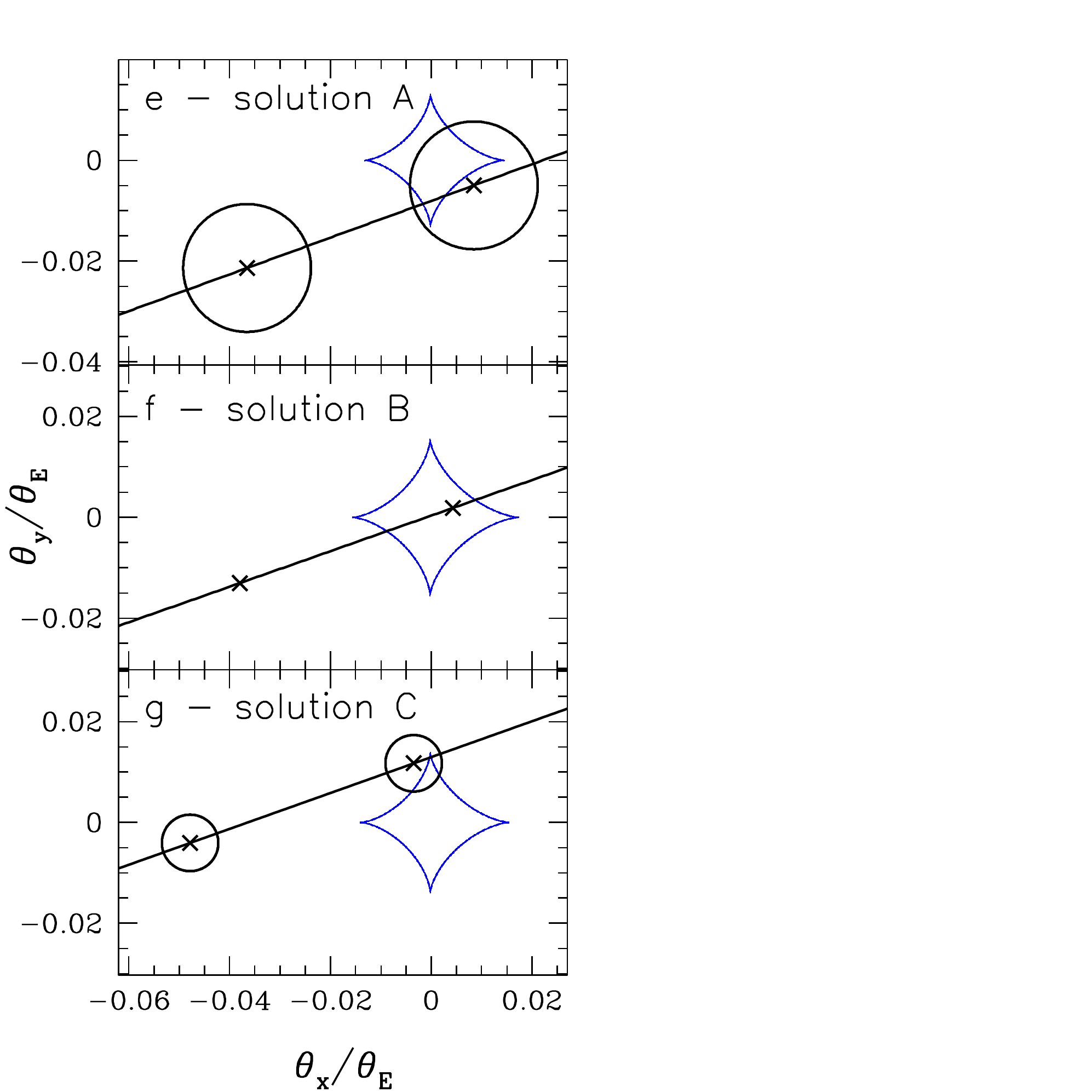}
\caption{Degenerate solutions for MOA-2012-BLG-006. 
{\it Top center panel d} presents anomaly part of light curves for three modes A, B, and C. 
All three light curves predict the second subevent at ${\rm HJD'}=6047$.  
{\it Remaining large panels a, b, and c} show the projection of the marginalized $\chi^2$ hypersurface. 
Red, orange, yellow, green, and blue points correspond to $\Delta\chi^2$ of 
$<1$, $<4$, $<9$, $<16$, and $<25$, respectively. 
Letters A, B, and C mark the three modes. 
We note that $u_{0,1}>0$ means that the companion and the host caustics are passed on opposite sides. 
The data presented on this plot are not the result of a single run, 
but a compilation of many simulations and were obtained using OGLE data only. 
Similar plot for OGLE and MOA data does not show 
significant differences except different levels of $\Delta\chi^2$. 
In particular, MOA data at ${\rm HJD'}=5965.2$ and $5967.2$ significantly contribute to preference of mode A over mode C. 
{\it The small panels on the right e, f, and g} show source trajectories of 
the three solutions relative to planetary caustic (blue curve). Crosses mark 
the source positions at epochs when OGLE data were taken. Circles have a radius of $\rho$. 
In the case of solution B, the source size of $\rho = 0.00016$ is too small to be seen. 
}
\label{fig:chi2}
\end{figure}

In the two-component parametrization, 
we can easily find and understand the very significant model degeneracies. 
We present the slice of $\chi^2$ hypersurface in Figure~\ref{fig:chi2}. 
There are three local minima for different source trajectories and $\rho$ values. 
The best-fitting model has a $\rho$ of $\approx0.012$ 
and the source trajectory that passes each caustic on a different side 
(see Figure~\ref{fig:lc}). 
The second best-fitting 
model has a much smaller source ($\rho\approx0.001$ or even smaller) and 
the source passing through the center of the planetary caustic. The small source
causes the characteristic U-shaped light curve 
(see the panel {\it d} of Figure~\ref{fig:chi2}) 
but both high-magnification parts of the light curve 
(when the source crosses the caustic) that reach $I$-band magnitude of $12.3$
are predicted to have 
occurred during the time when no data were taken. In this model, 
the brightest OGLE data point 
is taken close to the middle of the U-shaped trough. 
A Bayesian argument suggests that a model that predicts a large brightness 
variation during a time when no data were taken is a priori unlikely, 
although this argument alone cannot rule out such a model. 
However, we can exclude this model because it predicts 
unreasonably large relative lens-source geocentric proper motion 
$\mu=\theta_{\rm E}/t_{\rm E}=\theta_{\star}/(\rho t_{\rm E})$ on the order of 
$100~{\rm mas\,yr^{-1}}$, where $\theta_{\star}$ is the angular radius of the 
source star equal to $5.68 \pm 0.34~{\rm \mu as}$ (see Section~\ref{sec:source}). 
In the third solution, the source passes both caustics on the same side and has 
$\rho\approx0.006$. 
We reject this solution because it is worse 
than the first solution by  
$\Delta\chi^2 = 5.8$ ($\Delta\chi^2 = 51.3$ if MOA data are included).

Ultimately, we decided not to use the usual MCMC algorithm for fitting the model, 
but instead apply an alternative algorithm that is 
more suited to explore degenerate multidimensional distributions -- 
Multimodal Ellipsoidal Nested Sampling or MultiNest. 
The algorithm is described in 
detail and implemented by \citet{feroz08} and \citet{feroz09,feroz13}. 
In brief, MultiNest approximates a volume of parameter space for which 
$\chi^2$ is below some limiting value $\chi^2_0$ by a set of 
$N$ points. We used $N = 5000$ here for the final fitting, but 
reasonably good exploration of parameter space is achieved even for $N = 500$ 
when considering only OGLE data. 
Both parameter estimation and model selection 
result from a single simulation in which $\chi^2_0$ is reduced from one 
step to the next. In every step, one of the $N$ points with 
the highest $\chi^2$ is replaced 
by a point that has lower $\chi^2$ and was found by trial-and-error. 
The trial-and-error procedure randomly samples a union of ellipsoids 
enclosing the $N$ points according to the prior (which was uniform in linear parameters). 

For the final model fitting, we used both OGLE and MOA data and assumed 
$\rho > 0.007$. The $\rho$ constraint is equivalent to 
assuming that $\mu < 15~{\rm mas\,yr^{-1}}$. The Galactic simulation described 
below (with only $t_{\rm E}$ constrained) 
gives a probability of $\mu > 15~{\rm mas\,yr^{-1}}$ to be $<0.001$. 
We present the final model parameters obtained using MultiNest in Table~\ref{tab:params}. 
Even though the fitting was performed in the two-component parameterization, 
all the parameters in default parametrization except $\rho$ show 
symmetric posterior distributions. 
We note that the event MOA-2012-BLG-006 
was used in a statistical analysis of the exoplanet 
mass ratio function by \citet{suzuki16}. The parameters used there 
($t_{\rm E}=21.13~{\rm d}$, $u_0=1.3$, $q=0.01614$, $s=4.32$) 
slightly differ from those found in the present work. The \citet{suzuki16} analysis 
was performed independently from the present analysis.

\begin{table}
\caption{Double lens model parameters}             
\label{tab:params}      
\centering                        
\begin{tabular}{llr}        
\hline\hline                
Quantity & Unit & Value \\  
\hline                      
$t_0$ & &  $ 6046.87 \pm 0.12 $ \\
$u_0$ & &  $ 1.432 \pm 0.032 $ \\
$t_{\rm E}$ & d &  $ 20.69 \pm 0.35 $ \\
$\rho$ & &  $ 0.0119 _{- 0.0023 } ^{+ 0.0016 } $ \\
$\alpha$ & deg &  $ 20.17 \pm 0.20 $ \\
$s$ & &  $ 4.405 \pm 0.069 $ \\
$q$ & &  $ 0.01650 \pm 0.00055 $ \\
$F_s/F_{\rm base}$ & &  $ 0.981 \pm 0.054 $ \\
\hline
$\chi^2/{\rm dof}$  & & $2898.21/3130$ \\ 
\hline                                   
\end{tabular}
\tablefoot{The parameter $F_s/F_{\rm base}$ indicates ratio of source flux to baseline flux in the $I$-band. 
The value of $u_0$ is greater than one, hence, the host subevent would not 
normally be counted for the optical depth calculations. 
}
\end{table}

The problems with fitting the microlensing models described above are primarily 
caused by the poor coverage of the anomalous part of the light curve. 
Similar problems frequently appear during the analysis of poorly sampled anomalies. 
\citet{jaroszynski02} claimed that a single point anomaly in OGLE-2002-BLG-055 
could be explained by the planetary model. 
Later, \citet{gaudi04} showed that there is 
a plethora of models with non-planetary mass ratios that can fit the same 
light curve. Analysis of the planetary event MOA-2007-BLG-192 
revealed degenerate
cusp approach and caustic crossing solutions that could not be efficiently
sampled by MCMC runs \citep{bennett08} because 
of the huge number of steps needed to cross the $\chi^2$ barrier between them. 
We predict that MultiNest can solve the remaining problems in analysis of these, 
and other, poorly sampled events. 

Direct measurement of the lens mass, distance, and projected separation of the lens components 
requires microlensing parallax $\pi_{\rm E}$ \citep{gould00b} to be measured. 
We cannot measure or even put meaningful constraints on $\pi_{\rm E}$ for 
MOA-2012-BLG-006 because the host subevent is too short and the value of $u_0$ is  too large. 
The companion subevent could reveal the parallax signal only 
if it was sampled at a much higher cadence. 
Without a parallax measurement, we have to use source properties and Bayesian priors using a Galactic model 
to constrain the lens mass, distance, and projected separation of components. 

\subsection{Source properties} 
\label{sec:source}

\begin{figure}
\centering
\includegraphics[width=\hsize]{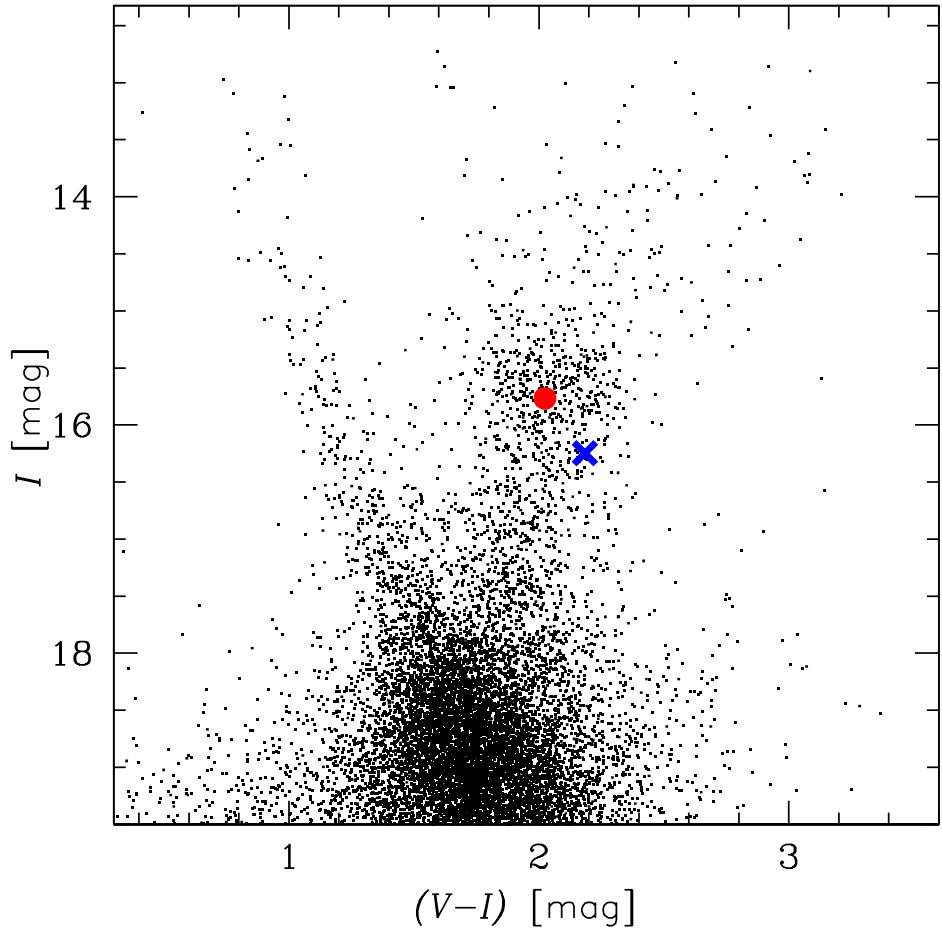}
\caption{Color-magnitude diagram for stars within $2\arcmin$ around MOA-2012-BLG-006. 
Red circle marks centroid of the red clump. 
The blue cross marks the position of the source. 
}
\label{fig:cmd}
\end{figure}

The lens mass and distance are crucial parameters for determining the nature of 
the lens system. These parameters cannot be directly derived from only the microlensing 
parameters like $t_{\rm E}$ and $\rho$. 
Microlensing events with finite source effects (and thus measured $\rho$) 
can only be used to estimate the physical properties of the lens if we can 
measure the angular Einstein ring radius $\theta_{\rm E}=\theta_{\star}/\rho$ 
and also measure the microlensing parallax or the lens flux. 
Here the estimate of $\theta_{\star}$ follows the method by \citet{yoo04b}. 
First, we construct the color-magnitude diagram for stars lying close to 
the event as presented in Figure~\ref{fig:cmd}. 
Second, we measure the properties of the red clump (RC): 
$I_{\rm RC}=15.767 \pm 0.017~{\rm mag}$ and
$(V-I)_{\rm RC}=2.024 \pm 0.007~{\rm mag}$. 
Third, by comparing these values with theoretical values, 
$I_{\rm RC,0}=14.381~{\rm mag}$ \citep[found by interpolation of Table~2 from][]{nataf13b} and
$(V-I)_{\rm RC,0}=1.06~{\rm mag}$ \citep{bensby11}, we find 
extinction $A_I = 1.386~{\rm mag}$ and
reddening $E(V-I)=0.964~{\rm mag}$. 
Fourth, we correct the source's unmagnified brightness 
($I_{s}=16.247~{\rm mag}$ and $V_{s}=18.390~{\rm mag}$) for extinction and obtain: 
$I_{s,0}=14.861~{\rm mag}$ and $V_{s,0}=16.040~{\rm mag}$. 
Fifth, the extinction-corrected brightness in visual bands is transformed 
to the near-infrared brightness of $K_{s,0}=13.314~{\rm mag}$ 
based on the intrinsic colors of giant stars \citep{bessell88}. 
Sixth, the angular source radius of $\theta_{\star} = 5.68 \pm 0.36~{\rm \mu as}$ is calculated 
using the relation between surface brightness and $(V-K)$ color by \citet{kervella04b}. 
This procedure results in $\theta_{\rm E} = \theta_{\star}/\rho = 0.489^{+0.126}_{-0.038}~{\rm mas}$. 

The event was also observed in $J$-band using adaptive optics (AO) at Keck telescope. 
\citet{bessell88} relations predict an extinction-free source brightness of 
$J_{s,0}=14.077 \pm 0.085~{\rm mag}$. The $J$-band extinction toward the event is 
$A_J = 0.58 \pm 0.16~{\rm mag}$ \citep{gonzalez12}. 
Hence, we predict observed source brightness of 
$J_s=14.66 \pm 0.19~{\rm mag}$.

\subsection{Galactic model} 
To derive the physical properties of the lens, we simulated microlensing events 
using the modified version of Galactic model by \citet{clanton14a} 
and we refer the reader to that 
paper for a detailed description. The model includes lenses from Galactic disc with 
a double-exponential density profile and boxy Gaussian bulge. 
The line-of-sight projected velocity of the Earth is calculated for a peak of the 
anomaly. 
The lens mass distribution is the same as in \citet{sumi11} model 1 limited to 
the main sequence lenses: power laws with $\alpha=1.3$ for $0.08\leq M/M_{\odot} < 0.7$  
and $\alpha=2.0$ for $0.7\leq M/M_{\odot} < 1.0$. 
Sources are placed at a distance of $7.8~{\rm kpc}$. 
The ensemble of simulated events is additionally weighted according to the measured
$t_{\rm E}$ and $\theta_{\rm E}$. 
No constraint on the lens flux was applied. 
The resulting distributions are used to estimate 
the physical properties of the lens. 
We used the Astropy package \citep{astropy13} to analyze the simulation. 

\begin{figure}
\centering
\includegraphics[width=\hsize]{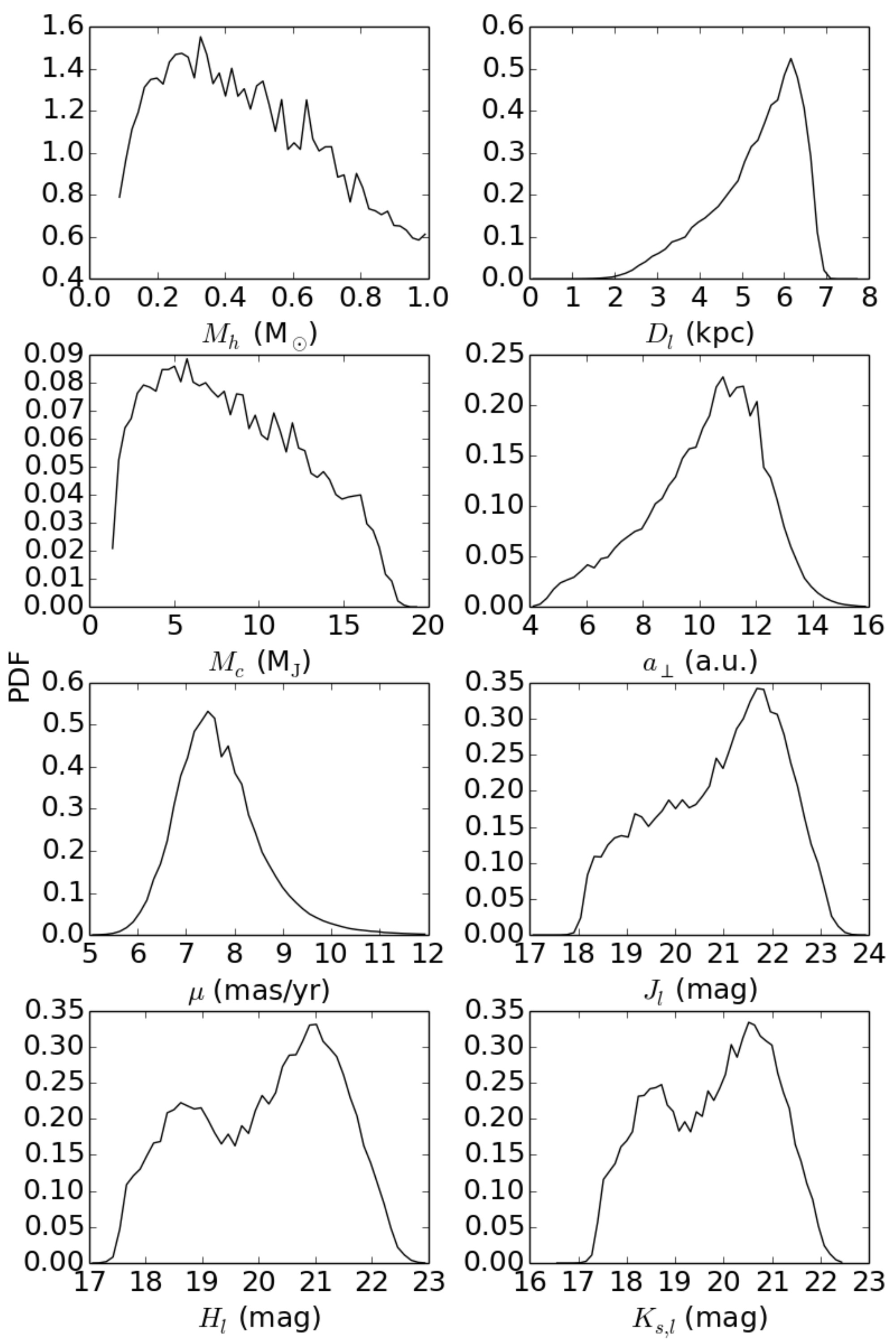}
\caption{Posterior distributions of parameters derived from the Galactic model.  
The three lower panels show predicted brightness of the lens in $J$, $H$, and $K_s$ filters (2MASS system). 
The extinction of $A_J = 0.58$, $A_H = 0.33$, and $A_{K_s} = 0.20$ \citep{gonzalez12} 
were assumed independent of the lens distance. 
}
\label{fig:gal}
\end{figure}

Figure~\ref{fig:gal} and Table~\ref{tab:gal} present 
the posterior distributions of the event parameters as 
derived from the Galactic model: 
$M_h$ and $M_c$ -- mass of the host and companion, respectively,  
$a_{\perp}$ -- projected separation of host and companion, 
$D_l$ -- distance to the lens, and 
$\mu$ -- relative proper motion of lens and source. 
We note that $a_{\perp}=10.2~{\rm a.u.}$ corresponds to deprojected 
semi-major axis (for a circular orbit with a random value of the cosine of the inclination) of 
$a = \sqrt{3/2}a_{\perp} = 12.5~{\rm a.u.}$ 
The probability that $M_c$ is above the frequently assumed minimum brown dwarf mass of $13M_J$ is $0.18$.
Figure~\ref{fig:gal} also includes predictions of lens near-infrared brightness 
based on \citet{dotter08} 6 Gyr isochrone for ${\rm [Fe/H]} = 0.0$ and $Y=0.27$. 

\begin{table}
\caption{Posterior statistics for event parameters using a Bayesian prior and a Galactic model. 
Mean values and $1-\sigma$ uncertainties are given.}
\label{tab:gal}      
\centering                  
\begin{tabular}{llr}        
\hline\hline                
Quantity & Unit & Value \\  
\hline                      
$M_h$ & ${\rm M_{\odot}}$ &  $ 0.49 _{- 0.23 } ^{+ 0.27 } $ \\
$M_c$ & ${\rm M_{J}}$ &  $ 8.4 _{- 3.9 } ^{+ 4.6 } $ \\
$D_l$ & ${\rm kpc}$ &  $ 5.3 _{- 1.3 } ^{+ 0.8 } $ \\
$a_{\perp}$ & ${\rm a.u.}$ &  $ 10.2 _{- 2.4 } ^{+ 1.8 } $ \\
$\mu$ & ${\rm mas/yr}$ &  $ 7.69 _{- 0.76 } ^{+ 1.1 } $ \\
\hline
\end{tabular}
\end{table}

\section{High-resolution observations}  

On July 18, 2013 ($1.2~{\rm yr}$ after peak of the event) we observed the microlensing event MOA-2012-BLG-006 with 
Near Infrared Camera 2 (NIRC-2) 
AO system mounted on the Keck-II telescope. We used the wide field ($40''\times40''$) camera with a pixel
scale of $0.04~{\rm arcsec}$ and $J$-band filter. We took four frames with an exposure time of $3 \times 10$ seconds at each of the
five dithered positions.  We corrected for dark and flat fields using standard procedure and stacked the
images using SWarp program from the Astr$O$matic suite of astronomy tools \citep{bertin10}.  The full width at half maximum
was $0.2~{\rm arcsec}$.  The aperture photometry was performed by running SExtractor \citep{bertin96} software.  
The photometric and astrometric calibration of the Keck 
images requires additional data and for this purpose we used VVV data. 
The VVV survey observed in $J$, $H$, and $K$ bands at the 4m VISTA telescope at Paranal Observatory (Chile). 
To process VVV images we followed the procedure described by \citet{beaulieu16} 
which includes calibration of photometry and astrometry to 2MASS system 
\citep{skurtskie06}. 

\begin{figure}
\centering
\includegraphics[width=.5\hsize]{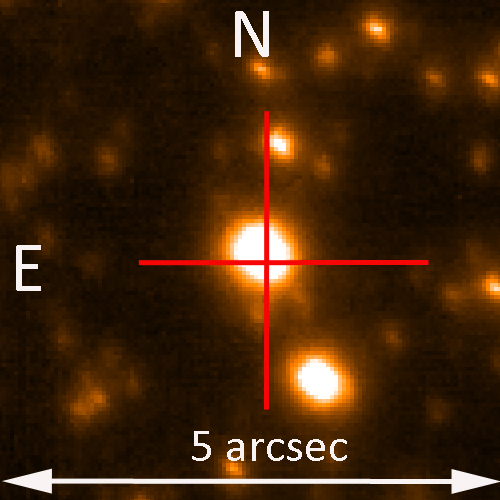}
\caption{Keck AO image of the event. The cross marks the expected position. 
}
\label{fig:ao}
\end{figure}

The source star is clearly identified on Keck image at the expected position (see Figure \ref{fig:ao}). 
We note that there is no significant blend at the sub-arcsec separation. 
The VVV brightness is 
$J_{VVV}=14.72 \pm 0.02$, 
$H_{VVV}=13.88 \pm 0.02$, and 
$K_{VVV}=13.62 \pm 0.03$. 
Based on cross-identification of the same stars in Keck and VVV data we estimate the error in 
absolute calibration of Keck photometry of $0.015~{\rm mag}$.  
The brightness measured on Keck image and calibrated to VVV data is $J_{\rm Keck}= 14.70\pm0.03$.

We can compare the total object brightness measured from the Keck image with the lens and source brightness 
estimated before. 
The fiducial lens mass from our Galactic model ($0.49~{\rm M_{\odot}}$) corresponds to absolute brightness 
on main sequence of $J_{l,0}=6.20$ \citep{dotter08}. 
The fiducial lens distance is $5.3~{\rm kpc}$, hence, 
it should be behind almost all extinction observed in this field. The expected brightness of the lens is 
hence $J_l = 20.40$ (see also Figure~\ref{fig:gal}). 
The optical data and VVV extinction predict the source brightness of $J_s=14.66 \pm 0.19$. 
Hence, the lens is on the order of $6~{\rm mag}$ fainter than the source and its contribution to the total 
light ($0.004~{\rm mag}$) is much smaller than the uncertainty in $J_s$. 
The object brightness measured on the Keck image $J_{\rm Keck}= 14.70 \pm 0.03$ is marginally brighter 
than the predicted source brightness $J_s=14.66 \pm 0.19$, that is, no light from the lens is detected. 
However, there is no statistical difference between 
the two measurements, primarily due to the large uncertainty of $J$-band extinction.

\section{Solving the mystery of OGLE-2002-BLG-045} 
\label{sec:ob02045}

\citet{skowron09} analyzed a number of microlensing events that were 
observed to repeat, that is, 
show more than one brightening episode. The second episode can be caused by 
either a companion to the lens or a companion to the source. The interesting 
finding by \citet{skowron09} was that 
the event OGLE-2002-BLG-045 showed two consecutive 
OGLE datapoints that are well separated from the main event and significantly 
brighter than the baseline. The two observations happened only four days apart 
(${\rm HJD'}=2455.7$ and $2459.6$) and were separated a few days from the previous 
(${\rm HJD'}=2448.7$) and the following (${\rm HJD'}=2463.6$) observations, which 
both were at the baseline. 
The short time between the two anomaly observations compared to $t_{\rm E} = 26.4~{\rm d}$ 
suggests that the anomaly could have been caused by 
a planetary companion to the lens ($q=0.008$ and $s=3.958$). As was pointed out by \citet{skowron09}, 
the only evidence for existence of the planet were the two data points brighter than the baseline, 
and therefore, the planet detection of the planetary companion was questionable. 
Because of this ambiguity, the putative planet OGLE-2002-BLG-045Lb
is not normally considered on the lists of the known microlensing planets 
\citep[e.g.,][]{zhu14,penny16a,mroz17a}. 

In order to verify the planetary signal in OGLE-2002-BLG-045 we  performed 
photometry of the archival data acquired by the previous phase of the MOA survey 
\citep[MOA-I;][]{yanagisawa00,bond01}. 
No signal of the planet was found. 
We also visually verified the OGLE images that resulted in two anomalous 
points and found that they were taken in non-photometric conditions. 
\citet{skowron09} inspected 4120 events, hence it is not surprising that 
in this sample they found an event with two consecutive erroneous measurements 
separated by a few $t_{\rm E}$ from the event peak.
We conclude that there is no convincing evidence that OGLE-2002-BLG-045L 
has a wide separation planet and the two data points brighter than baseline 
are simply observational artifacts.

\section{Conclusions} 

We presented the discovery of MOA-2012-BLG-006Lb -- an object  
a few times more massive than Jupiter, which can be classified
based on its mass either 
as a planet (most probable scenario) or a brown dwarf 
(if its mass is at the high end of the derived distribution). 
We detected microlensing signal not only due to this object but also due to its host star. 
The mass ratio is above $0.01$, that is, higher than the typical mass ratio 
of a protoplanetary disc to the parent star.
Hence, the lower-mass object could have formed independently 
and thus resembles brown dwarfs, even if its mass is smaller 
than the commonly assumed boundary of $13~{\rm M_{J}}$. 

The lack of parallax constraint precludes a direct 
measurement of the lens mass and projected separation of its components, 
but the large ratio between the two subevents' time separation 
and the event timescale suggests 
the companion is on a very wide orbit. 
Bayesian inference using a Galactic model results in a 
$0.49 \pm 0.25~{\rm M_{\odot}}$ host orbited by a $8.4 \pm 4.3~{\rm M_{J}}$ 
companion at a projected separation of 
$10.2\pm2.2~{\rm a.u.}$ 
The projected separation relative to Einstein ring radius of $s=4.4$ 
is the second largest among low-mass companions found by the microlensing 
technique after OGLE-2008-BLG-092 \citep[$s=5.3$;][]{poleski14c} and preceding 
MOA-2007-BLG-400 \citep[$s=2.9$ if wider of the two solutions is true;][]{dong09b}. 

The properties of MOA-2012-BLG-006Lb are similar to the small number of objects, 
either high-mass planets or brown dwarfs, that have been discovered around 
M stars via direct imaging and which typically have orbital separations of 
tens to hundreds of a.u. \citep[see e.g., Table 1 of ][]{lannier16}. 
The masses of such planets tend to be at least the same order of magnitude as 
the total amount of mass that comprised the protoplanetary disk within (and from) 
which we would expect them to have formed, presenting a challenge to our current 
understanding of giant planet formation. Nevertheless, 
the discovery of MOA-2012-BLG-006Lb suggests that whatever mechanisms are 
responsible for the formation of such objects, they seem to operate similarly 
in the immediate Solar neighborhood (where direct imaging finds them) 
and in other parts of the Galaxy, several kpc away 
(where only microlensing is sensitive to their detection).

The AO image of the event was taken using the Keck NIRC-2 camera. The contribution of 
the lens flux to the total observed flux could not be measured due to 
large uncertainty in extinction and 
the fact that the lens is expected to be substantially fainter than the source. 

We also showed that the anomaly observed in another event -- OGLE-2002-BLG-045 
-- is of instrumental origin. Hence, there is no planetary signature 
in that event.

\begin{acknowledgements}
The authors would like to thank Prof. A.~Gould for consultation. 
The OGLE project has received funding from the National Science Centre, Poland, 
grant MAESTRO 2014/14/A/ST9/00121 to A.U. 
OGLE Team acknowledges Profs. M.~Kubiak and G.~Pietrzy\'nski, former members of the team, 
for their contribution to the collection of the OGLE photometric data over the past years. 
J.P.B. and J.B.M. gratefully acknowledge support from ESO's DGDF 2014. 
C.C. was supported by an appointment to the NASA Postdoctoral Program at NASA Ames Research Center, 
administered by Universities Space Research Association under contract with NASA.
\end{acknowledgements}

\bibliographystyle{aa} 
\bibliography{paper}

\begin{thebibliography}{65}
\expandafter\ifx\csname natexlab\endcsname\relax\def\natexlab#1{#1}\fi

\bibitem[{{Alard}(2000)}]{alard00}
{Alard}, C. 2000, \aaps, 144, 363

\bibitem[{{Andrews} {et~al.}(2013){Andrews}, {Rosenfeld}, {Kraus}, \&
  {Wilner}}]{andrews13}
{Andrews}, S.~M., {Rosenfeld}, K.~A., {Kraus}, A.~L., \& {Wilner}, D.~J. 2013,
  \apj, 771, 129

\bibitem[{{Andrews} {et~al.}(2009){Andrews}, {Wilner}, {Hughes}, {Qi}, \&
  {Dullemond}}]{andrews09}
{Andrews}, S.~M., {Wilner}, D.~J., {Hughes}, A.~M., {Qi}, C., \& {Dullemond},
  C.~P. 2009, \apj, 700, 1502

\bibitem[{{Astropy Collaboration} {et~al.}(2013){Astropy Collaboration},
  {Robitaille}, {Tollerud}, {Greenfield}, {Droettboom}, {Bray}, {Aldcroft},
  {Davis}, {Ginsburg}, {Price-Whelan}, {Kerzendorf}, {Conley}, {Crighton},
  {Barbary}, {Muna}, {Ferguson}, {Grollier}, {Parikh}, {Nair}, {Unther},
  {Deil}, {Woillez}, {Conseil}, {Kramer}, {Turner}, {Singer}, {Fox}, {Weaver},
  {Zabalza}, {Edwards}, {Azalee Bostroem}, {Burke}, {Casey}, {Crawford},
  {Dencheva}, {Ely}, {Jenness}, {Labrie}, {Lim}, {Pierfederici}, {Pontzen},
  {Ptak}, {Refsdal}, {Servillat}, \& {Streicher}}]{astropy13}
{Astropy Collaboration}, {Robitaille}, T.~P., {Tollerud}, E.~J., {et~al.} 2013,
  \aap, 558, A33

\bibitem[{{Beaulieu} {et~al.}(2016){Beaulieu}, {Bennett}, {Batista}, {Fukui},
  {Marquette}, {Brillant}, {Cole}, {Rogers}, {Sumi}, {Abe}, {Bhattacharya},
  {Koshimoto}, {Suzuki}, {Tristram}, {Han}, {Gould}, {Pogge}, \&
  {Yee}}]{beaulieu16}
{Beaulieu}, J.-P., {Bennett}, D.~P., {Batista}, V., {et~al.} 2016, \apj, 824,
  83

\bibitem[{{Beichman} {et~al.}(2014){Beichman}, {Gelino}, {Kirkpatrick},
  {Cushing}, {Dodson-Robinson}, {Marley}, {Morley}, \& {Wright}}]{beichman14}
{Beichman}, C., {Gelino}, C.~R., {Kirkpatrick}, J.~D., {et~al.} 2014, \apj,
  783, 68

\bibitem[{{Bennett} {et~al.}(2008){Bennett}, {Bond}, {Udalski}, {Sumi}, {Abe},
  {Fukui}, {Furusawa}, {Hearnshaw}, {Holderness}, {Itow}, {Kamiya}, {Korpela},
  {Kilmartin}, {Lin}, {Ling}, {Masuda}, {Matsubara}, {Miyake}, {Muraki},
  {Nagaya}, {Okumura}, {Ohnishi}, {Perrott}, {Rattenbury}, {Sako}, {Saito},
  {Sato}, {Skuljan}, {Sullivan}, {Sweatman}, {Tristram}, {Yock}, {Kubiak},
  {Szyma{\'n}ski}, {Pietrzy{\'n}ski}, {Soszy{\'n}ski}, {Szewczyk},
  {Wyrzykowski}, {Ulaczyk}, {Batista}, {Beaulieu}, {Brillant}, {Cassan},
  {Fouqu{\'e}}, {Kervella}, {Kubas}, \& {Marquette}}]{bennett08}
{Bennett}, D.~P., {Bond}, I.~A., {Udalski}, A., {et~al.} 2008, \apj, 684, 663

\bibitem[{{Bensby} {et~al.}(2011){Bensby}, {Ad{\'e}n}, {Mel{\'e}ndez}, {Gould},
  {Feltzing}, {Asplund}, {Johnson}, {Lucatello}, {Yee}, {Ram{\'{\i}}rez},
  {Cohen}, {Thompson}, {Bond}, {Gal-Yam}, {Han}, {Sumi}, {Suzuki}, {Wada},
  {Miyake}, {Furusawa}, {Ohmori}, {Saito}, {Tristram}, \& {Bennett}}]{bensby11}
{Bensby}, T., {Ad{\'e}n}, D., {Mel{\'e}ndez}, J., {et~al.} 2011, \aap, 533,
  A134

\bibitem[{{Bertin}(2010)}]{bertin10}
{Bertin}, E. 2010, {SWarp: Resampling and Co-adding FITS Images Together},
  Astrophysics Source Code Library ascl.net/1010.068

\bibitem[{{Bertin} \& {Arnouts}(1996)}]{bertin96}
{Bertin}, E. \& {Arnouts}, S. 1996, \aaps, 117, 393

\bibitem[{{Bessell} \& {Brett}(1988)}]{bessell88}
{Bessell}, M.~S. \& {Brett}, J.~M. 1988, \pasp, 100, 1134

\bibitem[{{Bond} {et~al.}(2001){Bond}, {Abe}, {Dodd}, {Hearnshaw}, {Honda},
  {Jugaku}, {Kilmartin}, {Marles}, {Masuda}, {Matsubara}, {Muraki}, {Nakamura},
  {Nankivell}, {Noda}, {Noguchi}, {Ohnishi}, {Rattenbury}, {Reid}, {Saito},
  {Sato}, {Sekiguchi}, {Skuljan}, {Sullivan}, {Sumi}, {Takeuti}, {Watase},
  {Wilkinson}, {Yamada}, {Yanagisawa}, \& {Yock}}]{bond01}
{Bond}, I.~A., {Abe}, F., {Dodd}, R.~J., {et~al.} 2001, \mnras, 327, 868

\bibitem[{{Boss} {et~al.}(2003){Boss}, {Basri}, {Kumar}, {Liebert},
  {Mart{\'{\i}}n}, {Reipurth}, \& {Zinnecker}}]{boss03}
{Boss}, A.~P., {Basri}, G., {Kumar}, S.~S., {et~al.} 2003, Brown Dwarfs -- IAU
  Symposium, 211, 529

\bibitem[{{Cassan}(2008)}]{cassan08}
{Cassan}, A. 2008, \aap, 491, 587

\bibitem[{{Chabrier} {et~al.}(2014){Chabrier}, {Johansen}, {Janson}, \&
  {Rafikov}}]{chabrier14}
{Chabrier}, G., {Johansen}, A., {Janson}, M., \& {Rafikov}, R. 2014, Protostars
  and Planets VI, ArXiv e-prints 1401.7559, 619

\bibitem[{{Chauvin} {et~al.}(2015){Chauvin}, {Vigan}, {Bonnefoy}, {Desidera},
  {Bonavita}, {Mesa}, {Boccaletti}, {Buenzli}, {Carson}, {Delorme},
  {Hagelberg}, {Montagnier}, {Mordasini}, {Quanz}, {Segransan}, {Thalmann},
  {Beuzit}, {Biller}, {Covino}, {Feldt}, {Girard}, {Gratton}, {Henning},
  {Kasper}, {Lagrange}, {Messina}, {Meyer}, {Mouillet}, {Moutou}, {Reggiani},
  {Schlieder}, \& {Zurlo}}]{chauvin15}
{Chauvin}, G., {Vigan}, A., {Bonnefoy}, M., {et~al.} 2015, \aap, 573, A127

\bibitem[{{Clanton} \& {Gaudi}(2014)}]{clanton14a}
{Clanton}, C. \& {Gaudi}, B.~S. 2014, \apj, 791, 90

\bibitem[{{Claret} \& {Bloemen}(2011)}]{claret11}
{Claret}, A. \& {Bloemen}, S. 2011, \aap, 529, A75

\bibitem[{{Dodson-Robinson} {et~al.}(2009){Dodson-Robinson}, {Veras}, {Ford},
  \& {Beichman}}]{dodson-robinson09}
{Dodson-Robinson}, S.~E., {Veras}, D., {Ford}, E.~B., \& {Beichman}, C.~A.
  2009, \apj, 707, 79

\bibitem[{{Dong} {et~al.}(2009){Dong}, {Bond}, {Gould}, {Koz{\l}owski},
  {Miyake}, {Gaudi}, {Bennett}, {Abe}, {Gilmore}, {Fukui}, {Furusawa},
  {Hearnshaw}, {Itow}, {Kamiya}, {Kilmartin}, {Korpela}, {Lin}, {Ling},
  {Masuda}, {Matsubara}, {Muraki}, {Nagaya}, {Ohnishi}, {Okumura}, {Perrott},
  {Rattenbury}, {Saito}, {Sako}, {Sato}, {Skuljan}, {Sullivan}, {Sumi},
  {Sweatman}, {Tristram}, {Yock}, {MOA Collaboration}, {Bolt}, {Christie},
  {DePoy}, {Han}, {Janczak}, {Lee}, {Mallia}, {McCormick}, {Monard}, {Maury},
  {Natusch}, {Park}, {Pogge}, {Santallo}, {Stanek}, {{$\mu$}FUN Collaboration},
  {Udalski}, {Kubiak}, {Szyma{\'n}ski}, {Pietrzy{\'n}ski}, {Soszy{\'n}ski},
  {Szewczyk}, {Wyrzykowski}, {Ulaczyk}, \& {OGLE Collaboration}}]{dong09b}
{Dong}, S., {Bond}, I.~A., {Gould}, A., {et~al.} 2009, \apj, 698, 1826

\bibitem[{{Dong} {et~al.}(2007){Dong}, {Udalski}, {Gould}, {Reach}, {Christie},
  {Boden}, {Bennett}, {Fazio}, {Griest}, {Szyma{\'n}ski}, {Kubiak},
  {Soszy{\'n}ski}, {Pietrzy{\'n}ski}, {Szewczyk}, {Wyrzykowski}, {Ulaczyk},
  {Wieckowski}, {Paczy{\'n}ski}, {DePoy}, {Pogge}, {Preston}, {Thompson}, \&
  {Patten}}]{dong07}
{Dong}, S., {Udalski}, A., {Gould}, A., {et~al.} 2007, \apj, 664, 862

\bibitem[{{Dotter} {et~al.}(2008){Dotter}, {Chaboyer}, {Jevremovi{\'c}},
  {Kostov}, {Baron}, \& {Ferguson}}]{dotter08}
{Dotter}, A., {Chaboyer}, B., {Jevremovi{\'c}}, D., {et~al.} 2008, \apjs, 178,
  89

\bibitem[{{Feroz} \& {Hobson}(2008)}]{feroz08}
{Feroz}, F. \& {Hobson}, M.~P. 2008, \mnras, 384, 449

\bibitem[{{Feroz} {et~al.}(2009){Feroz}, {Hobson}, \& {Bridges}}]{feroz09}
{Feroz}, F., {Hobson}, M.~P., \& {Bridges}, M. 2009, \mnras, 398, 1601

\bibitem[{{Feroz} {et~al.}(2013){Feroz}, {Hobson}, {Cameron}, \&
  {Pettitt}}]{feroz13}
{Feroz}, F., {Hobson}, M.~P., {Cameron}, E., \& {Pettitt}, A.~N. 2013, ArXiv
  e-prints 1306.2144 [\eprint[arXiv]{1306.2144}]

\bibitem[{{Foreman-Mackey} {et~al.}(2016){Foreman-Mackey}, {Morton}, {Hogg},
  {Agol}, \& {Sch{\"o}lkopf}}]{foremanmackey16}
{Foreman-Mackey}, D., {Morton}, T.~D., {Hogg}, D.~W., {Agol}, E., \&
  {Sch{\"o}lkopf}, B. 2016, \aj, 152, 206

\bibitem[{{Gaudi}(1998)}]{gaudi98}
{Gaudi}, B.~S. 1998, \apj, 506, 533

\bibitem[{{Gaudi} \& {Han}(2004)}]{gaudi04}
{Gaudi}, B.~S. \& {Han}, C. 2004, \apj, 611, 528

\bibitem[{{Gonzalez} {et~al.}(2012){Gonzalez}, {Rejkuba}, {Zoccali}, {Valenti},
  {Minniti}, {Schultheis}, {Tobar}, \& {Chen}}]{gonzalez12}
{Gonzalez}, O.~A., {Rejkuba}, M., {Zoccali}, M., {et~al.} 2012, \aap, 543, A13

\bibitem[{{Gould}(2000)}]{gould00b}
{Gould}, A. 2000, \apj, 542, 785

\bibitem[{{Gould}(2008)}]{gould08}
{Gould}, A. 2008, \apj, 681, 1593

\bibitem[{{Gould} {et~al.}(2014){Gould}, {Udalski}, {Shin}, {Porritt},
  {Skowron}, {Han}, \& C.}]{gould14a}
{Gould}, A., {Udalski}, A., {Shin}, I.-G., {et~al.} 2014, Science, 345, 46

\bibitem[{{Grether} \& {Lineweaver}(2006)}]{grether06}
{Grether}, D. \& {Lineweaver}, C.~H. 2006, \apj, 640, 1051

\bibitem[{{Han}(2006)}]{han06}
{Han}, C. 2006, \apj, 638, 1080

\bibitem[{{Jaroszynski} \& {Paczynski}(2002)}]{jaroszynski02}
{Jaroszynski}, M. \& {Paczynski}, B. 2002, \actaa, 52, 361

\bibitem[{{Kains} {et~al.}(2012){Kains}, {Browne}, {Horne}, {Hundertmark}, \&
  {Cassan}}]{kains12}
{Kains}, N., {Browne}, P., {Horne}, K., {Hundertmark}, M., \& {Cassan}, A.
  2012, \mnras, 426, 2228

\bibitem[{{Kervella} {et~al.}(2004){Kervella}, {Bersier}, {Mourard},
  {Nardetto}, {Fouqu{\'e}}, \& {Coud{\'e} du Foresto}}]{kervella04b}
{Kervella}, P., {Bersier}, D., {Mourard}, D., {et~al.} 2004, \aap, 428, 587

\bibitem[{{Lannier} {et~al.}(2016){Lannier}, {Delorme}, {Lagrange}, {Borgniet},
  {Rameau}, {Schlieder}, {Gagn{\'e}}, {Bonavita}, {Malo}, {Chauvin},
  {Bonnefoy}, \& {Girard}}]{lannier16}
{Lannier}, J., {Delorme}, P., {Lagrange}, A.~M., {et~al.} 2016, \aap, 596, A83

\bibitem[{{Minniti} {et~al.}(2010){Minniti}, {Lucas}, {Emerson}, {Saito},
  {Hempel}, {Pietrukowicz}, {Ahumada}, {Alonso}, {Alonso-Garcia}, {Arias},
  {Bandyopadhyay}, {Barb{\'a}}, {Barbuy}, {Bedin}, {Bica}, {Borissova},
  {Bronfman}, {Carraro}, {Catelan}, {Clari{\'a}}, {Cross}, {de Grijs},
  {D{\'e}k{\'a}ny}, {Drew}, {Fari{\~n}a}, {Feinstein}, {Fern{\'a}ndez
  Laj{\'u}s}, {Gamen}, {Geisler}, {Gieren}, {Goldman}, {Gonzalez}, {Gunthardt},
  {Gurovich}, {Hambly}, {Irwin}, {Ivanov}, {Jord{\'a}n}, {Kerins}, {Kinemuchi},
  {Kurtev}, {L{\'o}pez-Corredoira}, {Maccarone}, {Masetti}, {Merlo},
  {Messineo}, {Mirabel}, {Monaco}, {Morelli}, {Padilla}, {Palma}, {Parisi},
  {Pignata}, {Rejkuba}, {Roman-Lopes}, {Sale}, {Schreiber}, {Schr{\"o}der},
  {Smith}, {}, {Soto}, {Tamura}, {Tappert}, {Thompson}, {Toledo}, {Zoccali}, \&
  {Pietrzynski}}]{minniti10}
{Minniti}, D., {Lucas}, P.~W., {Emerson}, J.~P., {et~al.} 2010, \na, 15, 433

\bibitem[{{Mr{\'o}z} {et~al.}(2017){Mr{\'o}z}, {Han}, {Udalski}, {Poleski},
  {Skowron}, {Szyma{\'n}ski}, {Soszy{\'n}ski}, {Pietrukowicz}, {Koz{\l}owski},
  {Ulaczyk}, {Wyrzykowski}, {Pawlak}, {Albrow}, {Cha}, {Chung}, {Jung}, {Kim},
  {Kim}, {Lee}, {Lee}, {Park}, {Pogge}, {Ryu}, {Shin}, {Yee}, {Zhu}, \&
  {Gould}}]{mroz17a}
{Mr{\'o}z}, P., {Han}, C., {Udalski}, A., {et~al.} 2017, \aj, 153, 143

\bibitem[{{Nataf} {et~al.}(2013){Nataf}, {Gould}, {Fouqu{\'e}}, {Gonzalez},
  {Johnson}, {Skowron}, {Udalski}, {Szyma{\'n}ski}, {Kubiak},
  {Pietrzy{\'n}ski}, {Soszy{\'n}ski}, {Ulaczyk}, {Wyrzykowski}, \&
  {Poleski}}]{nataf13b}
{Nataf}, D.~M., {Gould}, A., {Fouqu{\'e}}, P., {et~al.} 2013, \apj, 769, 88

\bibitem[{{Pejcha} \& {Heyrovsk{\'y}}(2009)}]{pejcha09}
{Pejcha}, O. \& {Heyrovsk{\'y}}, D. 2009, \apj, 690, 1772

\bibitem[{{Penny} {et~al.}(2016){Penny}, {Henderson}, \& {Clanton}}]{penny16a}
{Penny}, M.~T., {Henderson}, C.~B., \& {Clanton}, C. 2016, \apj, 830, 150

\bibitem[{{Poleski} {et~al.}(2014){Poleski}, {Skowron}, {Udalski}, {Han},
  {Koz{\l}owski}, {Wyrzykowski}, {Dong}, {Szyma{\'n}ski}, {Kubiak},
  {Pietrzy{\'n}ski}, {Soszy{\'n}ski}, {Ulaczyk}, {Pietrukowicz}, \&
  {Gould}}]{poleski14c}
{Poleski}, R., {Skowron}, J., {Udalski}, A., {et~al.} 2014, \apj, 795, 42

\bibitem[{{Ranc} {et~al.}(2015){Ranc}, {Cassan}, {Albrow}, {Kubas}, {Bond},
  {Batista}, {Beaulieu}, {Bennett}, {Dominik}, {Dong}, {Fouqu{\'e}}, {Gould},
  {Greenhill}, {J{\o}rgensen}, {Kains}, {Menzies}, {Sumi}, {Bachelet},
  {Coutures}, {Dieters}, {Dominis Prester}, {Donatowicz}, {Gaudi}, {Han},
  {Hundertmark}, {Horne}, {Kane}, {Lee}, {Marquette}, {Park}, {Pollard},
  {Sahu}, {Street}, {Tsapras}, {Wambsganss}, {Williams}, {Zub}, {Abe}, {Fukui},
  {Itow}, {Masuda}, {Matsubara}, {Muraki}, {Ohnishi}, {Rattenbury}, {Saito},
  {Sullivan}, {Sweatman}, {Tristram}, {Yock}, \& {Yonehara}}]{ranc16}
{Ranc}, C., {Cassan}, A., {Albrow}, M.~D., {et~al.} 2015, \aap, 580, A125

\bibitem[{{Sako} {et~al.}(2008){Sako}, {Sekiguchi}, {Sasaki}, {Okajima}, {Abe},
  {Bond}, {Hearnshaw}, {Itow}, {Kamiya}, {Kilmartin}, {Masuda}, {Matsubara},
  {Muraki}, {Rattenbury}, {Sullivan}, {Sumi}, {Tristram}, {Yanagisawa}, \&
  {Yock}}]{sako08}
{Sako}, T., {Sekiguchi}, T., {Sasaki}, M., {et~al.} 2008, Experimental
  Astronomy, 22, 51

\bibitem[{{Shvartzvald} {et~al.}(2014){Shvartzvald}, {Maoz}, {Kaspi}, {Sumi},
  {Udalski}, {Gould}, {Bennett}, {Han}, {Abe}, {Bond}, {Botzler}, {Freeman},
  {Fukui}, {Fukunaga}, {Itow}, {Koshimoto}, {Ling}, {Masuda}, {Matsubara},
  {Muraki}, {Namba}, {Ohnishi}, {Rattenbury}, {Saito}, {Sullivan}, {Sweatman},
  {Suzuki}, {Tristram}, {Wada}, {Yock}, {Skowron}, {Koz{\l}owski},
  {Szyma{\'n}ski}, {Kubiak}, {Pietrzy{\'n}ski}, {Soszy{\'n}ski}, {Ulaczyk},
  {Wyrzykowski}, {Poleski}, \& {Pietrukowicz}}]{shvartzvald14}
{Shvartzvald}, Y., {Maoz}, D., {Kaspi}, S., {et~al.} 2014, \mnras, 439, 604

\bibitem[{{Shvartzvald} {et~al.}(2016){Shvartzvald}, {Maoz}, {Udalski}, {Sumi},
  {Friedmann}, {Kaspi}, {Poleski}, {Szyma{\'n}ski}, {Skowron}, {Koz{\l}owski},
  {Wyrzykowski}, {Mr{\'o}z}, {Pietrukowicz}, {Pietrzy{\'n}ski},
  {Soszy{\'n}ski}, {Ulaczyk}, {Abe}, {Barry}, {Bennett}, {Bhattacharya},
  {Bond}, {Freeman}, {Inayama}, {Itow}, {Koshimoto}, {Ling}, {Masuda}, {Fukui},
  {Matsubara}, {Muraki}, {Ohnishi}, {Rattenbury}, {Saito}, {Sullivan},
  {Suzuki}, {Tristram}, {Wakiyama}, \& {Yonehara}}]{shvartzvald16a}
{Shvartzvald}, Y., {Maoz}, D., {Udalski}, A., {et~al.} 2016, \mnras, 457, 4089

\bibitem[{{Skowron} \& {Gould}(2012)}]{skowron12}
{Skowron}, J. \& {Gould}, A. 2012, ArXiv e-prints [\eprint[arXiv]{1203.1034}]

\bibitem[{{Skowron} {et~al.}(2011){Skowron}, {Udalski}, {Gould}, {Dong},
  {Monard}, {Han}, {Nelson}, {McCormick}, {Moorhouse}, {Thornley}, {Maury},
  {Bramich}, {Greenhill}, {Koz{\l}owski}, {Bond}, {Poleski}, {Wyrzykowski},
  {Ulaczyk}, {Kubiak}, {Szyma{\'n}ski}, {Pietrzy{\'n}ski}, {Soszy{\'n}ski},
  {OGLE Collaboration}, {Gaudi}, {Yee}, {Hung}, {Pogge}, {DePoy}, {Lee},
  {Park}, {Allen}, {Mallia}, {Drummond}, {Bolt}, {{$\mu$}FUN Collaboration},
  {Allan}, {Browne}, {Clay}, {Dominik}, {Fraser}, {Horne}, {Kains}, {Mottram},
  {Snodgrass}, {Steele}, {Street}, {Tsapras}, {RoboNet Collaboration}, {Abe},
  {Bennett}, {Botzler}, {Douchin}, {Freeman}, {Fukui}, {Furusawa}, {Hayashi},
  {Hearnshaw}, {Hosaka}, {Itow}, {Kamiya}, {Kilmartin}, {Korpela}, {Lin},
  {Ling}, {Makita}, {Masuda}, {Matsubara}, {Muraki}, {Nagayama}, {Miyake},
  {Nishimoto}, {Ohnishi}, {Perrott}, {Rattenbury}, {Saito}, {Skuljan},
  {Sullivan}, {Sumi}, {Suzuki}, {Sweatman}, {Tristram}, {Wada}, {Yock}, {MOA
  Collaboration}, {Beaulieu}, {Fouqu{\'e}}, {Albrow}, {Batista}, {Brillant},
  {Caldwell}, {Cassan}, {Cole}, {Cook}, {Coutures}, {Dieters}, {Dominis
  Prester}, {Donatowicz}, {Kane}, {Kubas}, {Marquette}, {Martin}, {Menzies},
  {Sahu}, {Wambsganss}, {Williams}, {Zub}, \& {PLANET
  Collaboration}}]{skowron11}
{Skowron}, J., {Udalski}, A., {Gould}, A., {et~al.} 2011, \apj, 738, 87

\bibitem[{{Skowron} {et~al.}(2016){Skowron}, {Udalski}, {Koz{\l}owski},
  {Szyma{\'n}ski}, {Mr{\'o}z}, {Wyrzykowski}, {Poleski}, {Pietrukowicz},
  {Ulaczyk}, {Pawlak}, \& {Soszy{\'n}ski}}]{skowron16a}
{Skowron}, J., {Udalski}, A., {Koz{\l}owski}, S., {et~al.} 2016, \actaa, 66, 1

\bibitem[{{Skowron} {et~al.}(2009){Skowron}, {Wyrzykowski}, {Mao}, \&
  {Jaroszy{\'n}ski}}]{skowron09}
{Skowron}, J., {Wyrzykowski}, {\L}., {Mao}, S., \& {Jaroszy{\'n}ski}, M. 2009,
  \mnras, 393, 999

\bibitem[{{Skrutskie} {et~al.}(2006){Skrutskie}, {Cutri}, {Stiening},
  {Weinberg}, {Schneider}, {Carpenter}, {Beichman}, {Capps}, {Chester},
  {Elias}, {Huchra}, {Liebert}, {Lonsdale}, {Monet}, {Price}, {Seitzer},
  {Jarrett}, {Kirkpatrick}, {Gizis}, {Howard}, {Evans}, {Fowler}, {Fullmer},
  {Hurt}, {Light}, {Kopan}, {Marsh}, {McCallon}, {Tam}, {Van Dyk}, \&
  {Wheelock}}]{skurtskie06}
{Skrutskie}, M.~F., {Cutri}, R.~M., {Stiening}, R., {et~al.} 2006, \aj, 131,
  1163

\bibitem[{{Spiegel} {et~al.}(2011){Spiegel}, {Burrows}, \&
  {Milsom}}]{spiegel11}
{Spiegel}, D.~S., {Burrows}, A., \& {Milsom}, J.~A. 2011, \apj, 727, 57

\bibitem[{{Street} {et~al.}(2013){Street}, {Choi}, {Tsapras}, {Han},
  {Furusawa}, {Hundertmark}, {Gould}, {Sumi}, {Bond}, {Wouters}, {Zellem},
  {Udalski}, {RoboNet Collaboration}, {Snodgrass}, {Horne}, {Dominik},
  {Browne}, {Kains}, {Bramich}, {Bajek}, {Steele}, {Ipatov}, {MOA
  Collaboration}, {Abe}, {Bennett}, {Botzler}, {Chote}, {Freeman}, {Fukui},
  {Harris}, {Itow}, {Ling}, {Masuda}, {Matsubara}, {Miyake}, {Muraki},
  {Nagayama}, {Nishimaya}, {Ohnishi}, {Rattenbury}, {Saito}, {Sullivan},
  {Suzuki}, {Sweatman}, {Tristram}, {Wada}, {Yock}, {OGLE Collaboration},
  {Szyma{\'n}ski}, {Kubiak}, {Pietrzy{\'n}ski}, {Soszy{\'n}ski}, {Poleski},
  {Ulaczyk}, {Wyrzykowski}, {{$\mu$}FUN Collaboration}, {Yee}, {Dong}, {Shin},
  {Lee}, {Skowron}, {De Almeida}, {DePoy}, {Gaudi}, {Hung}, {Jablonski},
  {Kaspi}, {Klein}, {Hwang}, {Koo}, {Maoz}, {Mu{\~n}oz}, {Pogge}, {Polishhook},
  {Shporer}, {McCormick}, {Christie}, {Natusch}, {Allen}, {Drummond},
  {Moorhouse}, {Thornley}, {Knowler}, {Bos}, {Bolt}, {PLANET Collaboration},
  {Beaulieu}, {Albrow}, {Batista}, {Brillant}, {Caldwell}, {Cassan}, {Cole},
  {Corrales}, {Coutures}, {Dieters}, {Dominis Prester}, {Donatowicz},
  {Fouqu{\'e}}, {Bachelet}, {Greenhill}, {Kane}, {Kubas}, {Marquette},
  {Martin}, {Menzies}, {Pollard}, {Sahu}, {Wambsganss}, {Williams}, {Zub},
  {MiNDSTEp}, {Alsubai}, {Bozza}, {Burgdorf}, {Calchi Novati}, {Dodds},
  {Dreizler}, {Finet}, {Gerner}, {Hardis}, {Harps{\o}e}, {Hessman}, {Hinse},
  {J{\o}rgensen}, {Kerins}, {Liebig}, {Mancini}, {Mathiasen}, {Penny}, {Proft},
  {Rahvar}, {Ricci}, {Scarpetta}, {Sch{\"a}fer}, {Sch{\"o}nebeck},
  {Southworth}, \& {Surdej}}]{street13}
{Street}, R.~A., {Choi}, J.-Y., {Tsapras}, Y., {et~al.} 2013, \apj, 763, 67

\bibitem[{{Sumi} {et~al.}(2010){Sumi}, {Bennett}, {Bond}, {Udalski}, {Batista},
  {Dominik}, {Fouqu{\'e}}, {Kubas}, {Gould}, {Macintosh}, {Cook}, {Dong},
  {Skuljan}, {Cassan}, {Abe}, {Botzler}, {Fukui}, {Furusawa}, {Hearnshaw},
  {Itow}, {Kamiya}, {Kilmartin}, {Korpela}, {Lin}, {Ling}, {Masuda},
  {Matsubara}, {Miyake}, {Muraki}, {Nagaya}, {Nagayama}, {Ohnishi}, {Okumura},
  {Perrott}, {Rattenbury}, {Saito}, {Sako}, {Sullivan}, {Sweatman}, {Tristram},
  {Yock}, {MOA Collaboration}, {Beaulieu}, {Cole}, {Coutures}, {Duran},
  {Greenhill}, {Jablonski}, {Marboeuf}, {Martioli}, {Pedretti}, {Pejcha},
  {Rojo}, {Albrow}, {Brillant}, {Bode}, {Bramich}, {Burgdorf}, {Caldwell},
  {Calitz}, {Corrales}, {Dieters}, {Dominis Prester}, {Donatowicz}, {Hill},
  {Hoffman}, {Horne}, {J{\o}rgensen}, {Kains}, {Kane}, {Marquette}, {Martin},
  {Meintjes}, {Menzies}, {Pollard}, {Sahu}, {Snodgrass}, {Steele}, {Street},
  {Tsapras}, {Wambsganss}, {Williams}, {Zub}, {PLANET Collaboration},
  {Szyma{\'n}ski}, {Kubiak}, {Pietrzy{\'n}ski}, {Soszy{\'n}ski}, {Szewczyk},
  {Wyrzykowski}, {Ulaczyk}, {OGLE Collaboration}, {Allen}, {Christie}, {DePoy},
  {Gaudi}, {Han}, {Janczak}, {Lee}, {McCormick}, {Mallia}, {Monard}, {Natusch},
  {Park}, {Pogge}, {Santallo}, \& {{$\mu$}FUN Collaboration}}]{sumi10}
{Sumi}, T., {Bennett}, D.~P., {Bond}, I.~A., {et~al.} 2010, \apj, 710, 1641

\bibitem[{{Sumi} {et~al.}(2011){Sumi}, {Kamiya}, {Bennett}, {Bond}, {Abe},
  {Botzler}, {Fukui}, {Furusawa}, {Hearnshaw}, {Itow}, {Kilmartin}, {Korpela},
  {Lin}, {Ling}, {Masuda}, {Matsubara}, {Miyake}, {Motomura}, {Muraki},
  {Nagaya}, {Nakamura}, {Ohnishi}, {Okumura}, {Perrott}, {Rattenbury}, {Saito},
  {Sako}, {Sullivan}, {Sweatman}, {Tristram}, {Udalski}, {Szyma{\'n}ski},
  {Kubiak}, {Pietrzy{\'n}ski}, {Poleski}, {Soszy{\'n}ski}, {Wyrzykowski},
  {Ulaczyk}, \& {Microlensing Observations in Astrophysics (MOA)
  Collaboration}}]{sumi11}
{Sumi}, T., {Kamiya}, K., {Bennett}, D.~P., {et~al.} 2011, \nat, 473, 349

\bibitem[{{Suzuki} {et~al.}(2016){Suzuki}, {Bennett}, {Sumi}, {Bond}, {Rogers},
  {Abe}, {Asakura}, {Bhattacharya}, {Donachie}, {Freeman}, {Fukui}, {Hirao},
  {Itow}, {Koshimoto}, {Li}, {Ling}, {Masuda}, {Matsubara}, {Muraki},
  {Nagakane}, {Onishi}, {Oyokawa}, {Rattenbury}, {Saito}, {Sharan}, {Shibai},
  {Sullivan}, {Tristram}, {Yonehara}, \& {MOA Collaboration}}]{suzuki16}
{Suzuki}, D., {Bennett}, D.~P., {Sumi}, T., {et~al.} 2016, \apj, 833, 145

\bibitem[{{Udalski}(2003)}]{udalski03}
{Udalski}, A. 2003, \actaa, 53, 291

\bibitem[{{Udalski} {et~al.}(2015){Udalski}, {Szyma{\'n}ski}, \&
  {Szyma{\'n}ski}}]{udalski15b}
{Udalski}, A., {Szyma{\'n}ski}, M.~K., \& {Szyma{\'n}ski}, G. 2015, \actaa, 65,
  1

\bibitem[{{Wilson} {et~al.}(2016){Wilson}, {H{\'e}brard}, {Santos}, {Sahlmann},
  {Montagnier}, {Astudillo-Defru}, {Boisse}, {Bouchy}, {Rey}, {Arnold},
  {Bonfils}, {Bourrier}, {Courcol}, {Deleuil}, {Delfosse}, F., {Ehrenreich},
  {Forveille}, {Moutou}, {Pepe}, {Santerne}, {S{\'e}gransan}, \&
  {Udry}}]{wilson16}
{Wilson}, P.~A., {H{\'e}brard}, G., {Santos}, N.~C., {et~al.} 2016, \aap, 588,
  A144

\bibitem[{{Wo\'zniak}(2000)}]{wozniak00}
{Wo\'zniak}, P.~R. 2000, \actaa, 50, 421

\bibitem[{{Yanagisawa} {et~al.}(2000){Yanagisawa}, {Muraki}, {Matsubara},
  {Abe}, {Masuda}, {Noda}, {Sumi}, {Kato}, {Fujimoto}, {Sato}, {Bond},
  {Rattenbury}, {Yock}, {Kilmartin}, {Hearnshaw}, {Reid}, {Sullivan}, {Carter},
  {Dodd}, {Nankivell}, {Rumsey}, {Honda}, {Sekiguchi}, {Yoshizawa}, {Nakamura},
  {Sato}, {Kabe}, {Kobayashi}, {Watase}, {Jugaku}, {Saito}, \&
  {Koribalsky}}]{yanagisawa00}
{Yanagisawa}, T., {Muraki}, Y., {Matsubara}, Y., {et~al.} 2000, Experimental
  Astronomy, 10, 519

\bibitem[{{Yoo} {et~al.}(2004){Yoo}, {DePoy}, {Gal-Yam}, {Gaudi}, {Gould},
  {Han}, {Lipkin}, {Maoz}, {Ofek}, {Park}, {Pogge}, {Mu-Fun Collaboration},
  {Udalski}, {Soszy{\'n}ski}, {Wyrzykowski}, {Kubiak}, {Szyma{\'n}ski},
  {Pietrzy{\'n}ski}, {Szewczyk}, {{\.Z}ebru{\'n}}, \& {OGLE
  Collaboration}}]{yoo04b}
{Yoo}, J., {DePoy}, D.~L., {Gal-Yam}, A., {et~al.} 2004, \apj, 603, 139

\bibitem[{{Zhu} {et~al.}(2014){Zhu}, {Penny}, {Mao}, {Gould}, \&
  {Gendron}}]{zhu14}
{Zhu}, W., {Penny}, M., {Mao}, S., {Gould}, A., \& {Gendron}, R. 2014, \apj,
  788, 73

\end{thebibliography}

\end{document}